\documentstyle[prl,aps,preprint,epsf]{revtex}
\input psfig
\psfull

\tightenlines
\def\etal{{\it et al.}}            
\begin{document}
\draft
\title{Version 0.0}
\title{Measurement of the form-factor ratios for
$D^+\rightarrow \overline K^{\,*0} e^+ \nu_e$}
\author{
    E.~M.~Aitala,$^9$
       S.~Amato,$^1$
    J.~C.~Anjos,$^1$
    J.~A.~Appel,$^5$
       D.~Ashery,$^{15}$
       S.~Banerjee,$^5$
       I.~Bediaga,$^1$
       G.~Blaylock,$^8$
    S.~B.~Bracker,$^{16}$
    P.~R.~Burchat,$^{14}$
    R.~A.~Burnstein,$^6$
       T.~Carter,$^5$
 H.~S.~Carvalho,$^{1}$
  N.~K.~Copty,$^{13}$
    L.~M.~Cremaldi,$^9$
 C.~Darling,$^{19}$
       K.~Denisenko,$^5$
       A.~Fernandez,$^{12}$
       G.~Fox,$^{13}$
       P.~Gagnon,$^2$
       K.~Gounder,$^9$
     A.~M.~Halling,$^5$
       G.~Herrera,$^4$
 G.~Hurvits,$^{15}$
       C.~James,$^5$
    P.~A.~Kasper,$^6$
       S.~Kwan,$^5$
    D.~C.~Langs,$^{11}$
       J.~Leslie,$^2$
       B.~Lundberg,$^5$
       S.~MayTal-Beck,$^{15}$
       B.~Meadows,$^3$
 J.~R.~T.~de~Mello~Neto,$^1$
    D.~Mihalcea,$^7$
    R.~H.~Milburn,$^{17}$
 J.~M.~de~Miranda,$^1$
       A.~Napier,$^{17}$
       A.~Nguyen,$^7$
  A.~B.~d'Oliveira,$^{3,12}$
       K.~O'Shaughnessy,$^2$
    K.~C.~Peng,$^6$
    L.~P.~Perera,$^3$
    M.~V.~Purohit,$^{13}$
       B.~Quinn,$^9$
       S.~Radeztsky,$^{18}$
       A.~Rafatian,$^9$
    N.~W.~Reay,$^7$
    J.~J.~Reidy,$^9$
    A.~C.~dos Reis,$^1$
    H.~A.~Rubin,$^6$
    D.~A.~Sanders,$^9$
 A.~K.~S.~Santha,$^3$
 A.~F.~S.~Santoro,$^1$
       A.~J.~Schwartz,$^{11}$
       M.~Sheaff,$^{4,18}$
    R.~A.~Sidwell,$^7$
    A.~J.~Slaughter,$^{19}$
    M.~D.~Sokoloff,$^3$
    J.~Solano,$^1$
       N.~R.~Stanton,$^7$
       K.~Stenson,$^{18}$
    D.~J.~Summers,$^9$
 S.~Takach,$^{19}$
       K.~Thorne,$^5$
    A.~K.~Tripathi,$^{10}$
       S.~Watanabe,$^{18}$
 R.~Weiss-Babai,$^{15}$
       J.~Wiener,$^{11}$
       N.~Witchey,$^7$
       E.~Wolin,$^{19}$
       D.~Yi,$^9$
       S.~Yoshida,$^{7}$                         
       R.~Zaliznyak,$^{14}$
       and
       C.~Zhang$^7$ \\
(Fermilab E791 Collaboration)
}

\address{
$^1$Centro Brasileiro de Pesquisas F{\'\i}sicas, Rio de Janeiro, Brazil \\
$^2$University of California, Santa Cruz, CA 95064\\
$^3$University of Cincinnati, Cincinnati, OH 45221 \\
$^4$CINVESTAV, Mexico\\
$^5$Fermilab, Batavia, IL 60510 \\
$^6$Illinois Institute of Technology, Chicago, IL 60616\\
$^7$Kansas State University, Manhattan, KS 66506 \\
$^8$University of Massachusetts, Amherst, MA 01003\\
$^9$University of Mississippi-Oxford, University, MS 38677 \\
$^{10}$The Ohio State University, Columbus, OH 43210\\
$^{11}$Princeton University, Princeton, NJ 08544\\
$^{12}$Universidad Autonoma de Puebla, Mexico \\
$^{13}$University of South Carolina, Columbia, SC 29208\\
$^{14}$Stanford University, Stanford, CA 94305 \\
$^{15}$Tel Aviv University, Tel Aviv, Israel\\
$^{16}$317 Belsize Drive, Toronto, Canada \\
$^{17}$Tufts University, Medford, MA 02155\\
$^{18}$University of Wisconsin, Madison, WI 53706 \\
$^{19}$Yale University, New Haven, CT 06511\\
}
\date{\today}
\maketitle
\begin{abstract}
We present a  measurement of the form-factor ratios 
$r_V=V(0)/A_1(0)$ and $r_2=A_2(0)/A_1(0)$ for the decay
$D^+\rightarrow \overline K^{\,*0} e^+ \nu_e$.
The measurement is based on a signal of approximately 3000 
$D^+\rightarrow \overline K^{\,*0} e^+ \nu_e,$ 
$\overline K^{\,*0}\rightarrow K^-\pi^+$ 
decays reconstructed
in data from charm hadroproduction experiment E791 at Fermilab.
The results are 
$r_V =1.84 \pm 0.11 \pm 0.08 $ and $r_2 =  0.71 \pm 0.08 \pm 0.09 $.
\end{abstract}

Semileptonic charm decays are useful for probing
the dynamics of hadronic currents because Cabibbo-Kobayashi-Maskawa
(CKM) weak mixing matrix elements for the charm sector are well known from 
unitarity constraints.  Form factors are Lorentz-invariant functions 
of $q^2$, the square of the virtual $W$ mass in the decay, that
describe how strong interactions modify the underlying weak 
decay \cite{ref:rmp}.  Form factor measurements in semileptonic
decays test a variety of models and nonperturbative
calculations.   
In addition, Heavy Quark Effective Theory\cite{ref:hqet} relates 
form factors in semileptonic charm decays to those in bottom decays 
(at the same four-velocity transfer), which 
are needed to extract the
weak mixing matrix elements $|V_{ub}|$ and $|V_{cb}|$ from semileptonic
bottom decays. The vector form factor $V(q^2)$ and
the axial-vector form factors $A_1(q^2)$ and $A_2(q^2)$ are
relevant to the decay $D^+\rightarrow \overline K^{\,*0} e^+ \nu_e$.  

Using data from charm hadroproduction experiment E791 at Fermilab, we
reconstruct about 3000 
$D^+\rightarrow \overline K^{\,*0} e^+ \nu_e$ (and charge-conjugate) decays 
(three times the largest previous sample~\cite{ref:e687})
and use the observed multidimensional distribution of
kinematic variables to extract the form factor ratios 
$r_V=V(0)/A_1(0)$ and $r_2=A_2(0)/A_1(0)$.  We assume the nearest-pole
dominance model for the $q^2$ dependence of the form factors,
$F(q^2) = F(0)/(1 - q^2/m^2_{pole})$ where $m_{pole}$ is
the appropriate vector or axial-vector pole mass: $m_V = 2.1$ GeV/$c^2$ or 
$m_A = 2.5$ GeV/$c^2$ \cite{ref:pdg96}.
     
E791 is a hadroproduction experiment \cite{ref:expt} that 
generated charm using a 500~GeV/$c$ $\pi^-$ beam incident on five thin targets
(one platinum, four diamond) separated by gaps of about 13.6 mm.
E791 ran with a loose transverse energy trigger and  
recorded $20 \times 10^9$ interactions 
during the 1991-92 Fermilab fixed-target run.
The important features of the E791 spectrometer for this
analysis are the tracking system (23 planes of silicon 
microstrip detectors, 45 planes of drift and 
proportional wire chambers, and two large-aperture dipole magnets), two 
threshold \v{C}erenkov counters 
that provide good $K/\pi$
separation over the momentum range 6 - 36~GeV/$c$,
and a lead--liquid-scintillator electromagnetic calorimeter.

In each event, we search for a candidate decay vertex
(secondary vertex) with 
unit charge made from three charged tracks, separated from the 
reconstructed production vertex
(primary vertex) by at least 15$\sigma_z$  
where $\sigma_z$ is the uncertainty in the longitudinal separation.  
The decay vertex is required to be at least one measurement error outside 
the nearest target material.
One charged particle must be consistent with being an electron
as determined by electromagnetic shower shape, the match between calorimeter
energy and tracking momentum, and the match between calorimeter and tracking
position measurements.
Electron identification efficiency is 
about 70\%, while the probability for a pion to be misidentified as 
an electron is only 1 - 2\%.  
One of the two remaining charged particles must have a \v{C}erenkov kaon 
signature.  Kaon identification efficiency is about 65\% in the 
momentum range 6 - 36~GeV/$c$ and lower above this range. 
The probability for misidentifying a pion as a kaon is about 5\% 
in the same momentum range and significantly lower above this
range.  Candidates consistent with being
misidentified $D^+\rightarrow K^-\pi^+\pi^+$ decays are removed.
If electron and kaon candidates are oppositely charged, the decay is
a candidate for $D^+\rightarrow \overline K^{\,*0} e^+ \nu_e,$ 
$\overline K^{\,*0}\rightarrow K^-\pi^+$.  We call them ``right-sign''
decays.  If electron and kaon candidates have the same charge, 
the decay is classified as ``wrong-sign''.  The 
wrong-sign sample is used to model background in the right-sign sample.
A clear excess of right-sign events compared to wrong-sign events 
is seen in the $K\pi$ invariant mass distribution 
at the $\overline K^{\,*0}$ mass.

\begin{figure}
\epsfxsize=3.0in
\hbox{\hfill\hskip+0.55in\epsffile{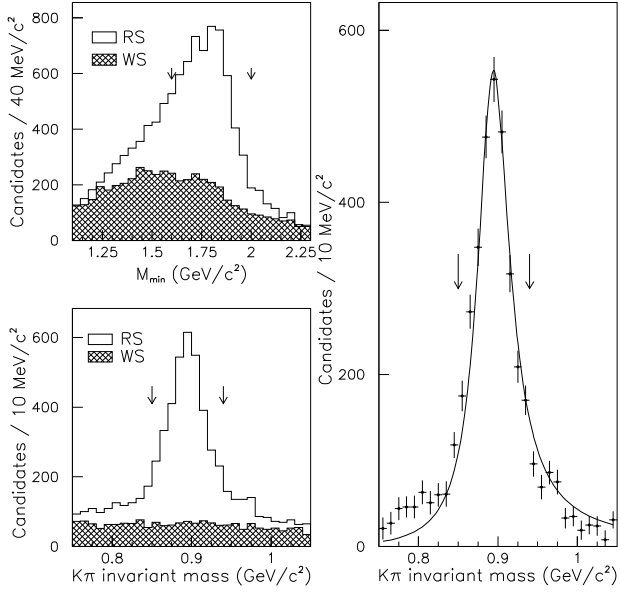}\hfill}
\caption{Final sample of 
$D^+\rightarrow \overline K^{\,*0} e^+ \nu_e$, 
$\overline K^{\,*0}\rightarrow K^-\pi^+$
candidates.  
The distributions are described in the text.
}
\label{fig:mass_plot}
\end{figure}

The final $K\pi e\nu_e$ sample (see Fig.~\ref{fig:mass_plot}) 
is optimized with a binary-decision-tree algorithm (CART\cite{ref:CART})
that finds the set of ``splits'' in a set of single 
parameters or linear combinations of parameters that best separates 
signal from background. 
We train CART using a subsample ($\approx 15\%$)
of the right-sign candidates for signal and the wrong-sign candidates 
for background.
CART selected a single cut involving a linear combination of 
four discrimination variables:  
(a)  separation significance of the candidate
decay vertex from target material, 
(b) distance of closest approach of the candidate $D$ momentum vector to 
the primary vertex allowing for the maximum kinematically-allowed
miss distance due to the unobserved neutrino,
(c) product over candidate $D$ decay tracks of the 
distance of closest approach of the track to
the secondary vertex divided by the distance of closest approach
to the primary vertex, where each distance is 
measured in units of measurement errors, and (d)  separation significance 
between the production and decay vertices. 
This selection criterion halved the number of 
wrong-sign events in the signal region, and kept 75\% 
of right-sign events.

Figure~\ref{fig:mass_plot} shows mass distributions for 
the final right-sign (RS) and 
wrong-sign (WS) $K \pi e \nu_e$ candidates.  The top left plot shows the 
distribution of $M_{min}$,
the minimum $K \pi e \nu_e$ mass kinematically allowed by the $D$ direction as
determined from the measured $K, \pi, \ $and $e$ momenta and the
positions of the primary and secondary vertices,
with a $0.75 < M(K \pi ) < 1.05$~GeV/$c^2$ cut
for both the right-sign and wrong-sign events.
The $M_{min}$ distribution for true $D^+\rightarrow K^-\pi^+ e^+ \nu_e$ events
(with no detector smearing) would have a cusp at the $D$ mass 
(1.869 GeV/$c^2$).
The bottom left plot in Fig.~\ref{fig:mass_plot} shows the 
$M(K\pi)$ distribution for events with 
$1.6 < M_{min} < 2.0$~GeV/$c^2$.
Of these, there are 3595 right-sign and 602 wrong-sign events  
with $0.85 < M(K\pi) < 0.94$~GeV/$c^2$
(indicated by the vertical arrows in the figures),  
which are used to extract the form-factor ratios.
The right-hand plot in Fig.~\ref{fig:mass_plot} shows the difference
between the right-sign and wrong-sign distributions.  The net
$K\pi$ signal is dominantly $K^{\,*0}$ as can be seen from
the superposed fit of the $M(K\pi)$ spectrum to a pure Breit-Wigner
shape with the mass and width fixed to the known values of the
$K^{\,*0}$ resonance.  There is an excess of events over that expected
for a Breit-Wigner distribution in the region below the $M(K\pi)$
range used in the analysis.  The assumption that wrong-sign events 
accurately model the size and shape of the right-sign background is 
addressed in the discussion of systematic uncertainties below. 

The kinematic variables used in extracting the form factor ratios
are the square of the invariant mass of the virtual $W$ ($q^2$) and 
three angles.
The polar angle $\theta_e$, measured in the virtual $W$ 
(or $e\nu$) rest frame, is the angle between the charged
lepton and the direction opposite the $K^{\,*0}$.
The polar angle $\theta_V$, measured in the $K^{\,*0}$ rest frame,
is the angle between the $K$ and the direction opposite
the virtual $W$.
The azimuthal angle $\chi$ is the angle between 
the momentum projections of the electron
and $K$ in the plane perpendicular to the $K^{\,*0}$ direction 
in the $D$ rest frame.
To calculate these variables, the neutrino momentum is estimated 
up to a two-fold ambiguity from the 
$D$ flight direction as determined by the measured
positions of the $D$ production  and decay points, and the 
measured momenta of the charged decay products.
Monte-Carlo simulation shows less 
kinematic variable smearing for the solution which results in
the lower $D$ momentum, so it is used.
 
We extract the form factors using an unbinned maximum-likelihood
method that uses a Monte-Carlo simulation in the evaluation of the likelihood
function \cite{ref:maxlik,ref:e691}.
The production physics and detector response are simulated for an event set
that is passed through the same analysis chain as the data.
The simulated events are generated with specified form-factor 
ratios (0.82 for $r_2$, 2.00 for $r_V$)
\cite{ref:dkrate}.
The likelihood of the data sample is calculated, for
any given set of theoretical parameters, by
computing the density of Monte-Carlo events in a specified volume around each
data point, where the simulated events are
distributed according to the theoretical parameters under consideration.
To avoid generating separate Monte-Carlo samples for
every set of theoretical parameters considered in the fit,
a single Monte-Carlo sample is reweighted so that the weighted 
events give the correct density about each data point.  As long as 
the Monte Carlo accurately
simulates both the charm production process and the detector response,
acceptance and smearing effects are automatically incorporated into the fit.
The wrong sign  candidates are used to incorporate backgrounds
into the fit with a similar technique.
We developed a second method to extract form factors that keeps both
neutrino momentum solutions and we use it as a check. To account for the
wrong solution, we use Monte Carlo simulation to determine a feedthrough
matrix that gives the probability that an event appearing in one region
of the space of measured kinematic variables would appear in another
region when the other solution for neutrino momentum is used. This matrix
and the observed distribution of data events (both solutions) determine
the fraction of data events that correspond to the correct
neutrino-momentum solution in each region of kinematic-variable space.
Each fraction is then used in a binned maximum likelihood fit, with
background modeled as in the first method.
                                                            
The first fitting technique gives 
$r_V = 1.84 \pm 0.11 \pm 0.08 $ and $r_2 =  0.71 \pm 0.08 \pm 0.09 $
with a correlation coefficient of $-0.13$.
The first error is statistical, and the second systematic (discussed below).
The above results have been corrected for biases of less than 10\% due to the
technique, determined from fitting Monte-Carlo samples
with known form-factor ratios.
The second fitting technique gives $r_V = 1.78 \pm 0.12$ and
$r_2 = 0.68 \pm 0.07$, where the errors are statistical only.
The Monte Carlo indicates no correction for bias is required.  
The results from the second fitting technique are consistent with
the first.

Possible sources of systematic uncertainties were considered 
and the most important are summarized 
in Table~\ref{tab:syserr}.  
To estimate the effects of possible 
inaccuracies in the Monte-Carlo simulation of the detector response,   
15 different sets of selection criteria were generated using different 
training samples for CART.  The spread in the resulting form-factor ratios 
gives our estimate of this systematic uncertainty\cite{ref:syserr}.    
We estimate the uncertainty due to our modeling of the background 
by varying our assumptions 
about the amount and the distribution of the background
in the four-dimensional kinematic variable space.  We have determined
that $D^{*+} \rightarrow D^0\pi^+, D^0 \rightarrow K^-e^+\nu_e$, and
$D^+ \rightarrow \overline K^{\,*0}\pi^0e^+\nu_e$ modes do not contribute
significantly to the background in the signal region.
Other sources of systematic uncertainty are the limited size of the
Monte-Carlo sample and uncertainties in charged-particle identification
efficiency.  The contributions from each source are added in quadrature
giving the total error of 0.08 and 0.09 for $r_V$ and $r_2$, respectively.  

\begin{figure}
\epsfxsize=3.0in
\hbox{\hfill\hskip+0.5in\epsffile{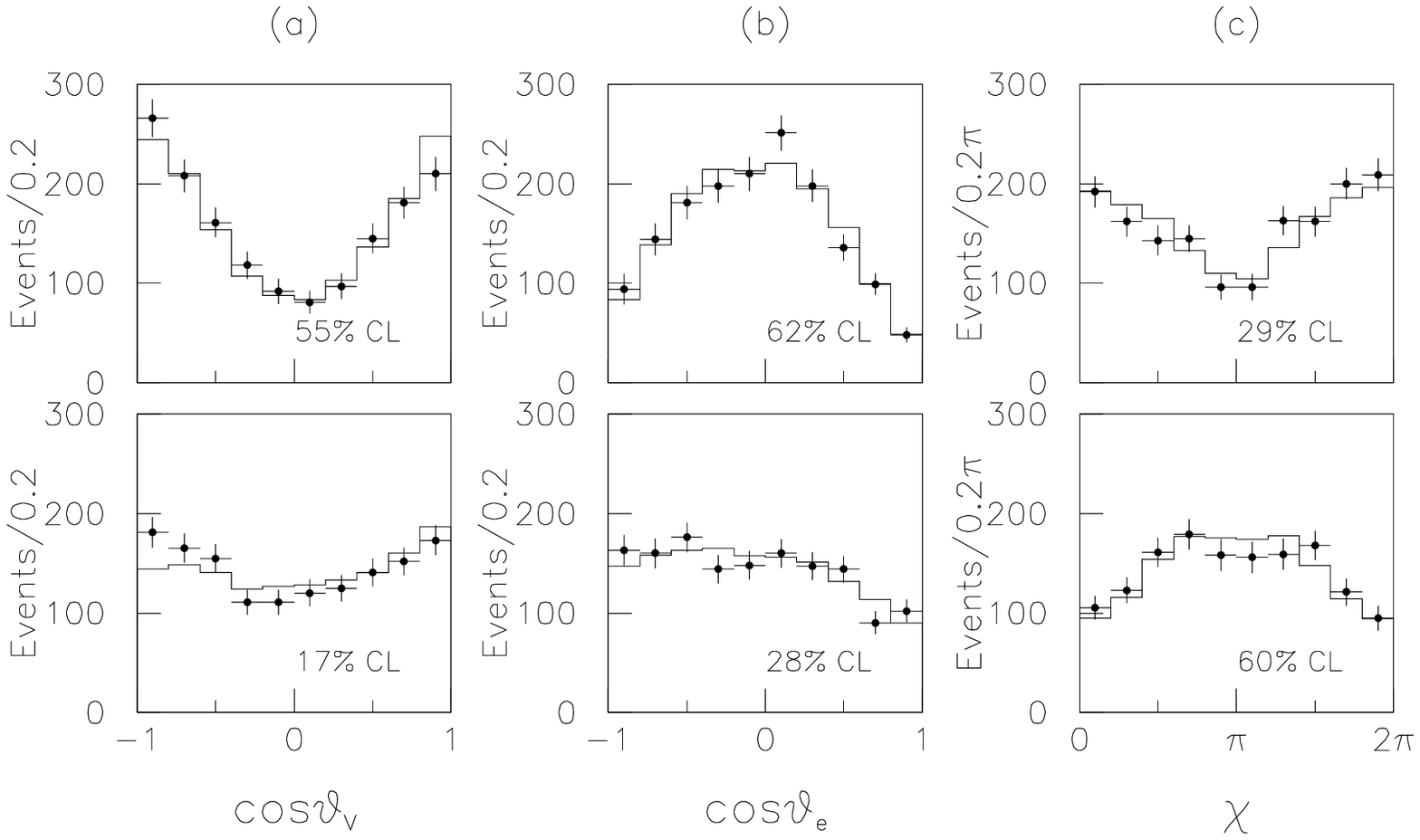}\hfill}
\caption{Data distributions (black dots with errors) overlaid with 
Monte Carlo (histogram) for 
(a) $\cos\theta_V$ for $q^2/q^2_{max} \leq 0.5$ (top)
and $q^2/q^2_{max} > 0.5$ (bottom),  
(b) $\cos\theta_e$ for $q^2/q^2_{max} \leq 0.5$ (top)
and $q^2/q^2_{max} > 0.5$ (bottom),
(c) $\chi$ for $\cos\theta_V \leq 0$ (top)
and $\cos\theta_V > 0$ (bottom).
The Monte-Carlo events are weighted
according to the best-fit values of the form-factor ratios.}
\label{fig:correl}
\end{figure}

Figure~\ref{fig:correl} shows the 
projections of the kinematic variables $\cos\theta_V$, $\cos\theta_e$,
and $\chi$ for 
data and for Monte-Carlo events that have been weighted according 
to the best fit values for the form-factor ratios.  To reveal the
correlations between the observed kinematic variables,
we show plots for each variable for two ranges of a second variable.
The confidence level for the consistency of the Monte Carlo and data
projections is shown on each plot. 

Throughout, we have assumed the nearest pole dominance $q^2$
dependence $F(q^2) = F(0)/(1 - q^2/m^2_{A,V})$ which can be approximated 
by a form linear in $q^2$, $F(q^2) = F(0)(1 + \rho^2 q^2)$.
This is a valid approximation in the accessible $q^2$ range
(0 - 0.947 GeV$^2$/$c^4$).  We perform a three-parameter fit 
for the slope $\rho^2_A$ 
and the form-factor ratios $r_V$ and $r_2$
(fixing $\rho_A^2/\rho_V^2 = m_V^2/m_A^2$).  The result for 
$r_V$ is $1.88 \pm 0.11$, which is close to 
the two-parameter
fit.  The results for $r_2$ and $\rho_A^2$ are  
strongly anticorrelated, so the statistical uncertainty on
these two parameters is large.  The result for $r_2$ 
is $0.98^{+0.14}_{-0.15}$, which is about two standard 
deviations higher than the result of the two-parameter fit.  
The result for $\rho^2_A$ is $-0.06^{+0.10}_{-0.09}$ GeV$^{-2}c^4$, 
which is about two standard deviations lower than the theoretical expectation 
$\rho^2_A = 1/m_A^2 = 0.16$ GeV$^{-2}c^4$.  

Table~\ref{tab:results} compares the form-factor ratios measured by this 
experiment and previously published results.  All measurements are in accord.
Using the world
averages for $B(D^+\rightarrow \overline{K}^{\,*0}e^+\nu_e)$ and the 
$D^+$ lifetime~\cite{ref:pdg96},
we extract the values of the form factors $A_1, A_2$, and $V$ 
at $q^2 = 0$ and at $q^2 = q^2_{max}$.  We account for both
the finite width of the $\overline K^{\,*0}$ and the correlation between the 
measured form-factor ratios~\cite{ref:ryd}. 
Table~\ref{tab:theory} compares
these results with predictions from Lattice 
Gauge calculations~\cite{ref:ape,ref:wupp,ref:ukqcd,ref:elc} and a
quark model calculation~\cite{ref:isgw} based on Heavy Quark Effective Theory.
The former are in fair agreement with the experimental results, 
while the latter is not, in particular for $A_2(q^2_{max})$. 

In summary, we have used a sample of 3000 signal events to 
extract the form-factor ratios in the
decay $D^+\rightarrow \overline{K}^{\,*0}e^+\nu_e$:
$r_V = 1.84 \pm 0.11 \pm 0.08$ and $r_2 = 0.71 \pm 0.08 \pm 0.09$.
The combined statistical and systematic uncertainties are a 
bit less than half 
the best previous measurement.  The form-factor
ratios are important for improving our understanding of the
dynamics of hadronic currents and might improve our ability to 
extract CKM matrix elements from $B$ semileptonic decays.

We thank the staffs of Fermilab and
all the participating institutions for assistance.  
This work was supported by the
Brazilian Conselho Nacional de Desenvolvimento Cient\'\i fico e
Technol\'{o}gico, CONACyT (Mexico), the Israeli Academy
of Sciences and Humanities, the U.S.\ Dept.\ of Energy, the U.S.-Israel
Binational Science Foundation, and the U.S. National Science Foundation.
Fermilab is operated by the Universities Research Association
under contract with the U.S.\ Dept.\ of Energy.

\begin{table}[t]
\begin{center}
\caption{Contributions to the systematic uncertainty.}
\label{tab:syserr}
\begin{tabular}{lcc}
Source of Uncertainty& $\sigma_{r_V}$ & $\sigma_{r_2}$ \\ \hline
Simulation of detector effects & 0.03 & 0.03 \\
Monte Carlo volume size & 0.04 & 0.04 \\
Background volume size & 0.04 & 0.05 \\
Amount of background & 0.02 & 0.05 \\
Particle identification efficiency& 0.05 & 0.01 \\
Fitting technique & 0.01 & 0.01 \\ \hline
Total Estimate & 0.08 & 0.09 \\
\end{tabular}
\end{center}
\end{table}

\begin{table}
\begin{center}
\caption{Results of this analysis (E791) and comparison with
other experiments.
The approximate number of signal events and 
the lepton type used in each analysis is also listed.}
\label{tab:results}
\begin{tabular}{lllr} 
Experiment & $r_V = V(0)/A_1(0)$         & $r_2 = A_2(0)/A_1(0)$ &
Events \\ \hline
E791 
 & $1.84 \pm 0.11 \pm 0.08$        & $0.71 \pm 0.08 \pm 0.09$ &
$ 3000$ ($e$) \\
E687\cite{ref:e687} & $1.74\pm0.27\pm0.28$              &
 $0.78\pm0.18\pm0.10$ &
$ 900$ ($\mu$) \\
E653\cite{ref:e653} & $2.00{^{+0.34}_{-0.32}}{\pm0.16}$ &
 $0.82{^{+0.22}_{-0.23}}\pm0.11$ &
$ 300$ ($\mu$) \\
E691\cite{ref:e691} & $2.0\pm0.6\pm0.3$                 & $0.0\pm0.5\pm0.2$ &
$ 200$ ($e$) \\
\end{tabular}
\end{center}
\end{table}

\begin{table}
\begin{center}
\caption{Form factors extracted from the measured ratios 
$r_V$ and $r_2$,
and several theoretical predictions.}
\label{tab:theory}
\renewcommand{\arraystretch}{1.15}
\begin{tabular}{llll}
Group & $A_1(0)$ & $A_2(0)$ & $V(0)$  \\ \hline   
E791 & $0.58\pm0.03$ & $0.41\pm0.06$ 
& $1.06\pm0.09$  \\    
APE~\cite{ref:ape} & $0.67\pm0.11$ & $0.49\pm0.34$ & $1.08\pm0.22$ \\    
Wuppertal~\cite{ref:wupp} & $0.61^{+0.11}_{-0.09}$ & $0.83^{+0.23}_{-0.22}$ 
& $1.34^{+0.31}_{-0.28}$ \\
UKQCD~\cite{ref:ukqcd} & $0.70^{+0.07}_{-0.10}$ & $0.66^{+0.10}_{-0.15}$ 
& $1.01^{+0.30}_{-0.13}$ \\  
ELC~\cite{ref:elc} & $0.64\pm0.16$ & $0.41\pm0.28$ & $0.86\pm0.24$ 
\\ \hline     
 & $A_1(q^2_{max})$ & $A_2(q^2_{max})$ & $V(q^2_{max})$ \\ \hline
E791 & $0.68\pm0.04$ & $0.48\pm0.08$ & $1.35\pm0.12$ \\
ISGW2~\cite{ref:isgw} & $0.70$ & $0.94$ & $1.52$ \\     
\end{tabular}  
\end{center}
\end{table}

\begin{thebibliography}{99}
\bibitem{ref:rmp} J.D.~Richman and P.R.~Burchat, 
Rev.~Mod.~Phys.~\textbf{67}, 893 (1995).
\bibitem{ref:hqet} N.~Isgur and M.B.~Wise, Phys.~Rev.~D \textbf{42}, 
2388 (1990).
\bibitem{ref:e687} P.L.~Frabetti \etal, Phys.~Lett.~B \textbf{307}, 262 (1993).
\bibitem{ref:pdg96}{\it Review of Particle Physics,} Phys.~Rev.~D \textbf{54}
(1996).
\bibitem{ref:expt} J.A.~Appel, Ann.~Rev.~Nucl.~Part.~Sci.~\textbf{42}, 
367 (1992); D.J.~Summers \etal, 
XXVII Rencontre de Moriond, 
Les Arcs, France (15-22 March 1992) 417.
\bibitem{ref:CART} L.~Brieman \etal,
{\it Classification and Regression Trees} (Chapman and Hall, New York, 1984).
\bibitem{ref:maxlik} D.M.~Schmidt, R.J.~Morrison, and M.S.~Witherell,
Nucl.\ Instrum.\ Methods A \textbf{328}, 547 (1993).
\bibitem{ref:e691} J.C.~Anjos \etal, Phys.~Rev.~Lett.~\textbf{65}, 2630 (1990).
\bibitem{ref:dkrate} For the general expression of the differential 
decay rate in terms of kinematic variables and form factors, see
Eqs.~113, 115, and 116 in \cite{ref:rmp}, or
Eqs.~2 and 3 on p.~482 in \cite{ref:pdg96}.
In ref.~\cite{ref:pdg96} $\cos\theta_e$ is defined as the angle between the 
charged lepton and the vector meson direction.
\bibitem{ref:syserr} We remove the contribution to the spread in
form-factor ratios due to the partial statistical independence of the 
15 samples, determined from Monte Carlo.
\bibitem{ref:e653} K.~Kodama \etal, Phys.~Lett.~B \textbf{274}, 246 (1992).
\bibitem{ref:ryd} A.~Ryd, private communication, using software originally
written by R.~Culbertson.
\bibitem{ref:ape} C.R.~Allton \etal, Phys.~Lett.~B \textbf{345}, 513 (1995).
\bibitem{ref:wupp} S.~Gusken \etal, Nucl.~Phys.~Proc.~Suppl. \textbf{47},
485 (1996).
\bibitem{ref:ukqcd} J.~Nieves \etal, Nucl.~Phys.~Proc.~Suppl. \textbf{42},
431 (1995).
\bibitem{ref:elc} A.~Abada \etal, Nucl.~Phys.~B \textbf{416}, 675 (1994).
\bibitem{ref:isgw} D.~Scora and N.~Isgur, Phys.~Rev.~D \textbf{52}, 
2783 (1995). 
\end{thebibliography}
\end{document}